\pdfoutput=1
\documentclass[12pt,fleqn]{article}

 \usepackage[frozencache=true,cachedir=minted-cache]{minted} 
\usepackage{arxiv}
\usepackage[utf8]{inputenc} % allow utf-8 input
\usepackage[T1]{fontenc}    % use 8-bit T1 fonts
\usepackage{url}            % simple URL typesetting
\usepackage{booktabs}       % professional-quality tables
\usepackage{amsfonts}       % blackboard math symbols
\usepackage{nicefrac}       % compact symbols for 1/2, etc.
\usepackage{microtype}      % microtypography

\usepackage{graphicx}
\usepackage[numbers]{natbib}
\usepackage{doi}
\usepackage{pgfplots}
\usepackage{listings}
\usepackage{algorithm}
\usepackage{algorithmicx,algpseudocode}
\usepackage[labelfont=bf, figurename=Fig.]{caption}
\usepackage{subcaption}
\usepackage{xcolor}
\usepackage{amsmath}
\usepackage{mathtools}
\usepackage{tabularx}
\usepackage{hyperref}       % hyperlinks

\lstdefinestyle{CStyle}{
    backgroundcolor=\color{backgroundColour},   
    commentstyle=\color{mGreen},
    keywordstyle=\color{magenta},
    numberstyle=\tiny\color{mGray},
    stringstyle=\color{mPurple},
    basicstyle=\footnotesize,
    breakatwhitespace=false,         
    breaklines=true,                 
    captionpos=b,                    
    keepspaces=true,                 
    numbers=left,                    
    numbersep=5pt,                  
    showspaces=false,                
    showstringspaces=false,
    showtabs=false,                  
    tabsize=2,
    language=C
}
\title {An AD based library for Efficient Hessian and Hessian-Vector Product Computation on GPU
}
\author{{\bf Desh Ranjan}\\
\textit{Department of Computer Science} \\
\textit{Old Dominion University}\\
\textit{Norfolk, Virginia, USA} \\
{\tt dranjan@cs.odu.edu}
\and
{\bf Mohammad Zubair}\\
\textit{Department of Computer Science} \\
\textit{Old Dominion University}\\
\textit{Norfolk, Virginia, USA} \\
{\tt zubair@cs.odu.edu}}
\renewcommand{\shorttitle}{Efficient Hessian and Hessian-Vector Product Computation on GPU}

\begin{document}
\maketitle
\begin{abstract}
The Hessian-vector product computation appears in many scientific applications such as in optimization and finite element modeling. Often there is a need for computing Hessian-vector products at many data points concurrently. We propose an automatic differentiation (AD) based method, CHESSFAD (Chunked HESSian using Forward-mode AD), that is designed with efficient parallel computation of Hessian and Hessian-Vector products in mind. CHESSFAD computes second-order derivatives using forward mode and exposes parallelism at different levels that can be exploited on accelerators such as NVIDIA GPUs. In CHESSFAD approach, the computation of a row of the Hessian matrix is independent of the computation of other rows. Hence rows of the Hessian matrix can be computed concurrently. The second level of parallelism is exposed because CHESSFAD approach partitions the computation of a Hessian row into chunks, where different chunks can be computed concurrently. CHESSFAD is implemented as a lightweight header-based C++ library that works both for CPUs and GPUs. We evaluate the performance of CHESSFAD for performing a large number of independent Hessian-Vector products on a set of standard test functions, and compare its performance to other existing header-based C++ libraries such as {\tt autodiff}. Our results show that CHESSFAD performs better than {\tt autodiff}, on all these functions with improvement ranging from 5-50\% on average. 
%on average $20\%$  for Rosebrock, $5\%$ for Ackley, and $49\%$ for Fletcher-Powell 
We also analyze its efficiency on GPUs as the number of variables in the function grows. 
We demonstrate that our approach is easily parallelizable and enables us to work with Hessian of a function of a large number of variables, which was not possible in sequential implementation. For example, the sequential execution time required for the Hessian-vector product for two variables is approximately enough to compute the Hessian-vector product for 
16 variables on GPU for all three functions.  A basic analysis of the number of arithmetic operations needed for computing the Hessian using the CHESSFAD approach is also provided. 
\end{abstract}

% \begin{IEEEkeywords}
% Automatic Differentiation, Hessian Computation, Hessian-vector product computation, GPU Computing.
% \end{IEEEkeywords}

\section{Introduction}
This paper presents an efficient automatic differentiation (AD) based method, CHESSFAD (Chunked HESSian using Forward-mode AD), for computing Hessian of a multivariate function. The key idea behind CHESSFAD is to extend the dual number approach for first-order AD in forward mode to compute the second-order derivatives so as to make it amenable to parallelism. We define a dual-number-like structure called $hDual$, which stores the value, the first-order derivatives, and the second-order derivatives of functions/expressions. Each real number is replaced by an $hDual$ in the function definition. We define overload operators on $hDual$s to correctly propagate the values of the function and the derivatives forward. A key idea in CHESSFAD is to allow for parameterized $hDual$s where each $hDual$ keeps only a certain number of first and second-order derivatives rather than all of them which could be prohibitive, especially for GPU implementations. This allows for computation of the Hessian in ``chunks" where a chunk is a contiguous set of second-order derivatives in a row of the Hessian. 
%These chunks can be computed in parallel and the implementation on GPUs can be optimized by making a proper choice of chunk size. We explain this in greater detail in later sections of the paper.
 
Several scientific applications require Hessian-vector products at many data points. These computations over multiple data points can be parallelized easily on emerging accelerators such as NVIDIA and INTEL GPUs. A single instance of Hessian computation can be further parallelized. Our approach to Hessian computation exposes parallelism at different levels. In our approach, the computation of a row of the Hessian matrix is independent of the computation of other rows. Hence rows of the Hessian matrix can be computed concurrently.  In the context of the Hessian-vector product, dot products associated with multiplying rows of the Hessian matrix with the vector can be executed concurrently.  The second level of parallelism is exposed in our approach because we partition the computation of a Hessian row into chunks, where different chunks can be computed concurrently. Keeping the chunk size small reduces the memory requirements but increases the number of arithmetic operations that need to be performed. Thus chunking allows for time-space tradeoff possibilities which allows for performance optimization on GPUs. In particular, since fast-access memory is at a premium in GPUs, keeping chunk size small might be beneficial. 

Another key feature of CHESSFAD is that it is implemented as a lightweight easy-to-use header-based C++ library. To use the library in CUDA C/C++ program or a C++ sequential program, one essentially needs to include it in the program and replace the variables of type {\tt double} with type $hDual$ in the function definition. This can be easily accomplished by writing a templated function on the data type. 

 \subsection { Related Work} Similar idea has been used in a previous paper to implement the computation of the Hessian of a function \cite{SzirmayKalos2020HigherOA}. In that implementation, one keeps a vector of size $(n+1)(n+2)/2$ for each real number. The first component of the vector keeps the function value, the next $n$ components keep the first-order derivatives and the remaining $n(n+1)/2$ components keep the upper half of the Hessian. By properly defining the rules for overload operators one can propagate the correct values of all these components. We explain this in some detail later in the paper. {\tt autodiff} is another popular software to calculate higher-order derivatives that allows for computation of up to fourth order derivatives using forward-mode AD \cite{autodiff}.  
 %MATLAB provides a method for computing the Hessian using a similar forward-mode AD idea but also computes it only one second-order derivative at a time {\CITATION needed}. 
 In our approach, we provide an elegant and efficient way of parameterizing the number of second-order derivatives computed at a time. As far as we know, none of these methods has a simple header-based C++ library implementation for computing Hessian-Vector products on GPUs. 
 
 Numerous other libraries provide the ability to compute Hessian-Vector products using AD-like JAX\cite{jax2018github}, ADEPT\cite{adept} and HAD\cite{li2019}. JAX is a popular package that allows for efficient computation of Hessian-Vector products using both forward and reverse mode automatic differentiation. Computing Hessian-Vector Product using Reverse-mode AD could potentially be faster due to better asymptotic complexity. HAD is an efficient simple reverse-mode AD-based lightweight header-based library. Once again, we are unaware of simple header-based C++ implementations to compute the Hessian-Vector product for these libraries.
 
Recently, several AD tools have been developed for GPU architectures, for example, Refs.~\cite{enzyme2021,sacado2022,Schule2022, Ifrim_2023,Bl_hdorn_2021}. Enzyme~\cite{enzyme2021} performs AD of GPU kernels using an LLVM-based plugin~\cite{Schule2022} that can generate kernel gradients in CUDA or ROCm. In Ref.~\cite{enzyme2021}, the authors demonstrated that the AD performance on a set of benchmarks is within an order of magnitude of the performance of the source program. Sacado~\cite{sacado2022} implements forward-mode AD using operator overloading with expression templates in C++.
The GPU support for Sacado is accomplished through Kokkos~\cite{Kokkos2022}, a C++ portability framework that works with GPUs of different vendors.

 We evaluated our approach on a set of functions that have been utilized in previous efforts \cite{Fletcher1963ARC}. Each of the functions chosen can be defined generically for increasing number of variables. These functions represent a variety of optimization problems, an area where the Hessian-Vector product is heavily utilized. We first showed that our base (sequential) implementation is reasonably efficient by comparing it with two other C++ header-based libraries -- one based on forward-mode AD ({\tt autodiff}) and another based on reverse-mode AD (HAD). We then explored the performance of CHESSFAD on GPUS for these functions as compared to our sequential implementation and showed that it is possible to scale up the performance to allow for its applications to solve reasonably large problems. While the current implementation of CHESSFAD is still less than fully optimized the results obtained are promising.

 The paper is organized as follows. Section~\ref{sec:AD} briefly discusses the basics of AD and implementation of forward mode AD using dual numbers. Section~\ref{sec:Extending} presents our approach to extending this to computing second-order derivatives and in particular the Hessian-vector product. Section~\ref{sec:CHESSFAD-Implementation} provides some details of the library implementation including the data structures used and code snippets to illustrate the implementation of overload operators. 
 In Section~\ref{sec:Analysis} we carry out a basic analysis that relates the number of arithmetic operations needed to chunk size. 
 Section~\ref{sec:HVP-Implementation} summarizes the key ideas and some details of the GPU implementation of CHESSFAD. Section~\ref{sec:Experimental-Results} presents our experimentation details and the results of the experimentation together with some discussion of the observed results. We conclude with some comments and future research directions in Section~\ref{sec:Conclusion}.

\section{Automatic Differentiation}
\label{sec:AD}
Most real-valued mathematical functions used in scientific and engineering applications are compositions of simple functions. Hence, the derivative of an output function can be computed by repeatedly using the chain rule. For a composite function $ f(g(x), h(x))$, the derivative, ${\frac{d f}{d x}}$, is computed as follows.

$${\frac{d f}{d x}} = {\frac{\partial f}{\partial g}} \times {\frac{d g}{d x}} + {\frac{\partial f}{\partial h}} \times {\frac{d h}{d x}}$$

 Different AD methods can be interpreted as a chain rule executed either in \textit{forward mode} or \textit{reverse mode}~\cite{griewank_book}. In forward mode, the chain rule is executed from right to left, meaning that derivatives of all intermediate quantities with respect to a particular input parameter are computed. In reverse mode, the chain rule is executed from left to right, meaning that all derivatives of a particular output function with respect to all intermediate quantities are computed. Forward mode is often easier to implement and has a smaller memory footprint. The complexity of an efficient forward-mode AD implementation is proportional to the number of input parameters. 
 \iffalse 
 One forward-mode AD method that is widely used in CFD and multidisciplinary optimization is based on complex-number theory~\cite{NewmanAndersonWhitefield1998,LynessMoler1967,Lyness1967}. Many forward-mode AD implementations use the dynamic approach, for example, Refs.~\cite{CppAD,Naumann2015,Walther09,Bischof2002}.
 Reverse-mode AD has complexity proportional to the number of output functions and is suitable for applications where there are many input parameters and few output functions. Reverse mode is more challenging to implement and parallelize. Its implementation often requires additional memory and/or computations~\cite{Griewank2000,Hogan2014,Bucker2005ABo,Huckelheim2019,Tapenade2013}. Depending on the underlying GPU architecture and the application, the registers required for a reverse-mode implementation may greatly exceed the available registers on a device, resulting in spills and significant performance slowdown. In Ref.~\cite{Huck2020}, the authors demonstrated that an efficient forward-mode AD implementation outperforms the most efficient reverse-mode AD implementation for a function with several hundred input parameters. Note that in such a case, forward mode performs significantly more floating-point operations than reverse mode; the superior performance is achieved due to the reduction of memory-related overhead\cite{Huck2020,evels2016forwardmode}.
 \fi
 
 \subsection {Forward-mode AD using Simple and Multivariate Duals} We next review AD implementations based on simple and multivariate dual numbers. 
 \iffalse A brief analysis of operation count and memory requirements associated with such implementations is also presented.
 \fi

%\subsection{Simple Dual Numbers}

A simple dual number (or dual) is an ordered pair of real numbers. A dual-number forward-mode AD can be implemented as follows. All real numbers involved in function evaluation are replaced by duals. For each quantity $u$, the dual corresponding to $u$ stores the value of $u$ and the value of $\frac{d u}{d x}$. The mathematical operations are extended to duals. For example, for duals $u=\langle a_0,b_0 \rangle$ and $v=\langle a_1,b_1 \rangle$, we can define the operations $+, \times, {\mbox and} {\tt sin}$ \iffalse and ${\tt sqrt}$ \fi as below:

\begin{center}
$u + v = \langle a_0+a_1, b_0+b_1 \rangle$ \\
$u \times v = \langle a_0*a_1, a_0*b_1 + b_0*a_1 \rangle$ \\
${\tt sin}(u) = \langle {\tt sin}(a_0), {\tt cos}(a_0)*b_0 \rangle$ \\
%${\tt sqrt}(u) = \langle {\tt sqrt}(a_0), (0.5/{\tt sqrt}%(a_0))*b_0 \rangle$
\end{center}

Recursively performing the operations, one can carry forward both the real quantity in the first component and its derivative in the second component. With proper initialization and by repeating the computation of the function one can calculate all the individual derivatives of a multivariate function.
\iffalse The differentiated functions can be multivariate, i.e., $f= f(x_1,\ldots, x_n)$. Forward mode considers derivatives with respect to a single independent input parameter $x_k$. By convention, $\frac{d x_j} {d x_k} = 0$ if $j\neq k$ and $\frac{d x_k} {d x_k} = 1$. Hence, to compute the derivative $\frac{d f} {d x_k} $  at $x_1=r_1,\ldots, x_n=r_n$, we initialize the corresponding duals to $\langle r_1,0 \rangle, \langle r_2,0 \rangle, \ldots, \langle r_k,1 \rangle, \ldots, \langle r_n,0\rangle$. To obtain the derivatives with respect to several or all input parameters, one can repeat the computation with each desired value of $k$. The only thing that changes is the initialization of the duals.
\fi

A major advantage of the dual-number AD implementation is that it requires only minor modification of the source code. One can define a new type for duals with corresponding dual operators as a standalone library. Once this is done, one can change all real variables in the function calculation to duals. The operator overloading mechanism provided by languages like C++ automatically applies the dual operators where appropriate.

%\subsection{Multivariate Dual Numbers}\label{multivariate}
Calculation of multiple derivatives of a multivariate function using simple duals as described above incurs a penalty of repeated computations, e.g., those performed for computing the function itself. One way to alleviate this penalty is to compute several derivatives simultaneously. This can be done by using {\em multivariate dual numbers} (or multivariate duals). A multivariate dual of dimension $n$ is a vector with $n+1$ components. The first component of the vector stores the quantity itself, the other $n$ components store the derivatives of the quantity with respect to specified input parameters. Once again, with proper use of the overload operators the correct values of multiple derivatives can be carried forward.

Use of multivariate duals can reduce the operation count signifiantly. However this comes at 
%A significant reduction of the total operation count achieved %by using multivariate duals comes at 
a cost of increased memory use, which can have a negative impact on performance. 

% For example, if the function $f$ uses memory to store $m$ real variables, an implementation using simple duals doubles the memory requirement to $2m$. An implementation based on $n$-dimensional duals, increases the memory requirement to $(n+1)m$, which can be detrimental, or even prohibitive, for performance. To optimize performance for a specific application on a specific architecture, one can choose a suitable dimension of multivariate duals. For example, the AD implementation based on multivariate duals of dimension $\frac{n}{2}$ calls the function twice calculating $\frac{n}{2}$ derivatives each time. In this case, some computations repeat but the memory requirements are reduced to $(\frac{n}{2} +1)m$. Such a trade-off can improve AD performance~\cite{revels2016forwardmode}.
% \subsection{Maintaining the Integrity of the Specifications}

% The IEEEtran class file is used to format your paper and style the text. All margins, 
% column widths, line spaces, and text fonts are prescribed; please do not 
% alter them. You may note peculiarities. For example, the head margin
% measures proportionately more than is customary. This measurement 
% and others are deliberate, using specifications that anticipate your paper 
% as one part of the entire proceedings, and not as an independent document. 
% Please do not revise any of the current designations.

\section{Extending the dual number idea to compute second-order derivatives}
\label{sec:Extending}
Akin to use of dual numbers for automatic differentiation in forward mode, one can use dual number like structure to compute the second-order derivatives. This idea has been used in a previous paper to implement the computation of the Hessian of a function using idea akin to AD. In that implementation one keeps a vector of size $(n+1)(n+2)/2$ for each real number. The first component of the vector keeps the function value, the next $n$ components keep the first order derivatives and the remaining $n(n+1)/2$ components keep the upper half of the Hessian. By properly defining the rules for overload operators one can propagate the correct values of all these components.  However, keeping all first and second order derivatives can be highly prohibitive especially for GPU implementations even for relatively small values of $n$ because of space limitations (MORE PRECISE WORDING NEEDED). We present an approach that allows one to store only a few derivatives at a time and trade space for arithmetic operations in an easy and systematic way. We explain this in some detail below.    

\subsection{CHESSFAD Approach} Let us first imagine that we are interested in computing a particular second order-derivative ${\frac{\partial^2 f}{\partial x_i \partial x_j}}$ of an $n$-variate function $f(x_1, \ldots, x_n)$. This can be done by using a dual number like structure with four components for each real number. We call such a structure $hDual$. The first component of the $hDual$ keeps the function value, the second component keeps the partial derivative with respect to $x_i$, the third the partial derivative with respect to $x_j$ and the fourth the value of the second order-derivative with respect to $x_i$ and $x_j$ ( $\langle f, \frac{\partial f}{\partial x_i}, \frac{\partial f}{\partial x_j}, \frac{\partial^2 f} {\partial x_i \partial x_j} \rangle$). We can define operations on these $hDual$s just like dual numbers to properly propagate the correct values based on calculus rules for first and second order derivatives. We present some example calculus rules and corresponding operations below:

\begin{center}
$\frac{\partial{(u+v)}}{\partial x}= \frac{\partial u}{\partial x} + \frac{\partial u}{\partial x}  $\\
$\frac {\partial {uv}} {\partial x}  =  u \frac {\partial v} {\partial x} + v \frac {\partial u} {\partial x} $\\
$\frac{\partial {\tt sin} (u)} {\partial x}  = {\tt cos} (u) \frac {\partial u}{\partial x} $\\
\end{center}

 \begin{center}
 
$\frac{\partial^2 (u+v)}{{\partial x \partial y}}  = \frac {\partial^2 u} {{\partial x \partial y}} + \frac{\partial^2 v}{{\partial x \partial y}} $\\
$\frac{\partial^2 uv} {\partial x \partial y} = u\frac{\partial^2 v}{\partial x \partial y} + \frac {\partial u} {\partial x}\frac {\partial v} {\partial y} + \frac {\partial v}{\partial x}\frac {\partial u} {\partial y} + v\frac{\partial^2 u}{\partial x \partial y}  $\\
$\frac{\partial^2 {{\tt sin} (u)}} {\partial x \partial y}  = {\tt cos} (u) \frac {\partial^2 u} {\partial x \partial y} - {\tt sin} (u) \frac{\partial u}{\partial x}\frac{\partial u}{\partial y} $\\
\end{center}

We can then define the overloaded operators +, $\times$, and {\tt sin} on $hDual$s. Let $u = \langle u_0,u_1,u_2,u_3 \rangle$ and $v = (v_0,v_1,v_2,v_3)$ be two $hDual$s. Then the definitions below will propagate the correct values forward.

\begin{center}
$u+v = \langle u_0+v_0, u_1+v_1, u_2+v_2, u_3+v_3 \rangle $\\
$u \times v = \langle u_0*v_0, u_0*v_1 + v_0*u_1, u_0*v_2 +v_0*u_2, u_0*v_3 + u_1*v_2 + u_2*v_1 + u_3*v_0 \rangle$\\
${\tt sin} (u) = \langle {\tt sin}(u_0), {\tt cos}(u_0)*u_1, {\tt cos}(u_0)*u_2, {\tt cos}(u_0)*u_3 - {\tt sin}(u_0)*u_1*u_2\rangle$\\
%${\tt sqrt}(u) = \langle{\tt sqrt}(a_0), (0.5/{\tt sqrt}(a_0))*b_0, (0.5/{\tt %sqrt}(a_0)*c_0 \rangle$\\
\end{center}

Now if we wanted to calculate the second-order derivative ${\frac{\partial^2 f}{\partial x_i \partial x_j}}$ of an $n$-variate function $f(x_1,..x_n)$ on a point $\langle a_1,a_2,\ldots,a_n \rangle$ this could be done by evaluating the function on right set of $hDual$s with the overloaded operators. For each variable $x_k$, we create an $hDual$ $y$[$k$]. To compute the second order derivative $\frac{\partial^2 f}{\partial x_i \partial x_j}$ we should set the variables $y[1],\ldots y[k]$ as follows (see Algorithm~\ref{algo:initialize}):

% {THIS SHOULD BE CHANGED TO FIGURE}
% \begin{verbatim}
% procedure INITIALIZE(y,r,i,j)
% 1 for k=1 to n 
%     y_k[0] = r_k
% 2 for k= 1 to n 
%     for t = 1 to 3
%         y_k[t] = 0.0; 
% 3. y_i[1] = 1.0;
% 4. y_j[2] = 1.0;   
% \end{verbatim}

\begin{algorithm}[ht]
\small
\caption{\textsc{INITIALIZE}($n$, $y$[$n$], $a$[$n$], $i$, $j$)}
\label{algo:initialize}
\begin{algorithmic}[1]
\For {$k$ $\leftarrow$ 0 \textbf{to} $n-1$}
\State{$y$[$k$].$v$[$0$] $\leftarrow$ $a$[$k$]}
\For{$l \leftarrow 1$ {\mbox {\bf to}} $3$}
\State{$y$[$k$].$v$[$l$] $\leftarrow 0.0 $}
\EndFor
\If{($k = i$)}
\State{$y$[$k$].$v$[$1$] $\leftarrow 1.0$ }
\EndIf
%\State{$y$[$k$].$v$[$2$] $\leftarrow 0.0 $}
\If{ ($k = j$)}
\State{$y$[$k$].$v$[$2$] $\leftarrow 1.0$ }
\EndIf
% \For {$ l \leftarrow 2$ \textbf{to} $\mathrm{3}$}
%\State{$y$[$k$].$v$[3] $\leftarrow 0.0$ }
\EndFor
\end{algorithmic}
\end{algorithm}

Now evaluating $f$ on $\langle y$[$1$],$\ldots y$[$n$]$\rangle$ would result in an $hDual$ whose first component would be $f$, the second component $\frac {\partial f} {\partial x_i}$ , the third component $\frac {\partial f} {\partial x_j}$ and the fourth component $\frac{\partial^2 f}{\partial x_i \partial x_j}$  where all these values are all computed at the point $\langle a_1,..a_n \rangle$. One can actually use this idea to compute the complete Hessian of $f$ by repeatedly computing the individual second order derivatives by properly initializing the $y[k]$ values and evaluating the function $f$ on $\langle y[1],y[2]...,y[n] \rangle$. The algorithm for doing so is presented as Algorithm ~2.

% {THIS SHOULD BE CHANGED TO FIGURE}
% \begin{verbatim}
% 1  for i = 1 to n 
% 2      for j=i to n
%        //initialize y to compute second derivative wrt x_i,x_j
% 3           INITIALIZE(y,r,i,j)
% 9           temp = f(y_1,y_2...,y_n)
% 10          H[i][j] = H[j,i] = temp[3];
% \end{verbatim}

\begin{algorithm}[ht]
\small
\caption{\textsc{HESSIAN}($f$, $n$, $a$[$n$], $H$[$n$][$n$])}
\label{algo:hessian-vector}
\begin{algorithmic}[1]
\State {$hDual$ $y$[$n$], $temp$}
% \State {$nchunk \leftarrow n/csize$}
\For {$ i \leftarrow 0$ \textbf{to} $n-1$}
\For {$ j \leftarrow 0$ \textbf{to} $n-1$}
% \State {$cstart \leftarrow  j} * csize $} 
\State {INITIALIZE($n$, $y$, $a$, $i$, $j$)}
\State {$temp \leftarrow f \langle hDual \rangle (y)$}

% \For {$ l \leftarrow 0$ \textbf{to} $\mathrm{csize}-1$}
\State {${H[i][j]} \leftarrow  temp.v[3] $}
% \EndFor
\EndFor
\EndFor
\end{algorithmic}
\end{algorithm}

Note that this way of computing the Hessian requires the computation of the function $f$ on $hDual$s $n^2$ times. This can be reduced to $n(n+1)/2$ using the symmetry of the Hessian. Algorithm {\textsc SYM-HESS} presented here accomplishes that. 

\begin{algorithm}[ht]
\small
\caption{\textsc{SYM-HESSIAN}($f$, $n$, $a$[$n$], $H$[$n$][$n$])}
\label{algo:symmetrichessian}
\begin{algorithmic}[1]
\State {$hDual$ $y$[$n$], $temp$}
% \State {$nchunk \leftarrow n/csize$}
\For {$ i \leftarrow 0$ \textbf{to} $n-1$}
\For {$ j \leftarrow i$ \textbf{to} $n-1$}
% \State {$cstart \leftarrow  j} * csize $} 
\State {INITIALIZE($n$, $y$, $a$, $i$, $j$)}
\State {$temp \leftarrow f \langle hDual \rangle (y)$}

% \For {$ l \leftarrow 0$ \textbf{to} $\mathrm{csize}-1$}
\State {${H[i][j]} \leftarrow  temp.v[3] $}
\State {$H[j][i] \leftarrow H[i][j]$}
% \EndFor
\EndFor
\EndFor
\end{algorithmic}
\end{algorithm}
This is how the Hessian is essentially computed using AD both in {\tt autodiff} and using MATLAB. Also note that in this method, we need to create an $hDual$ with $4$ components for each real-valued variable used to compute $f$. In contrast, if we were to keep vectors of size $(n+1)(n+2)/2$ for each real-valued variable as described in {\cite{SzirmayKalos2020HigherOA}, we will require a single computation of $f$. Depending on the value of $n$, this space requirement could be prohibitive. Moreover, each individual operation will require greater time. 

\subsection{Trading Space for Arithmetic Operations in Hessian and Hessian-Vector Product Computation}

The choices mentioned in the last paragraph of previous subsections are not the only two possibilities for computing the Hessian using the forward-mode AD idea. Another possibility is to compute the Hessian row-by-row one row at a time. In this case, when computing the second-order derivatives for row $i$, one needs to keep only the function value, the first-order derivatives $x_1,\ldots x_n$ and second-order derivatives w.r.t to $x_ix_1, x_ix_2,\ldots x_ix_n$ to propagate the values forward by proper definition of overload operators. Each $hDual$ now needs $2*n+1$ components but the function $f$ needs to be called $n$ times. This is a trade-off between the size of $hDual$s and the number of arithmetic operations needed to compute the Hessian. This can be further generalized by computing contiguous blocks or "chunks" of rows of Hessian at a time rather than the whole row. 
CHESSFAD does exactly that. It uses a parameter $csize$ that specifies the chunk size or how many second order-derivatives within a row would be computed at a time. Usually one would choose a value of $csize$ that exactly divides $n$. With the choice of $csize$, $hDual$s keeps an array of 2*($csize$+1) real numbers for each real number in the original function. The first component is the value, the second component is the first order derivative w.r.t. $x_i$, the next $csize$ components are the first order derivatives and the last $csize$ components are second-order derivatives ($\langle f, \frac{\partial f}{\partial x_i}, \frac{\partial f}{\partial x_j}, \ldots \frac{\partial f}{\partial x_{j+csize -1}}, \frac{\partial^2 f}{\partial x_i \partial x_j} \ldots \frac{\partial^2 f}{\partial x_i \partial x_{j+csize -1}} \rangle$). Each call to function $f$ computes $csize$ second order derivatives. The original function has to be called $n^2/csize$ times to compute the full Hessian. Using the symmetry of the Hessian, it is possible to compute it but avoid computing all the chunks which lie completely under the diagonal. The resulting algorithm is presented as Algorithm SCHUNK-HESS. A simple analysis of the time-space tradeoff when using the chunked Hessian is presented in section V. 

\begin{algorithm}[ht]
\small
\caption{\textsc{CHUNK-INIT}($n$, $y$[$n$], $a$[$n$], $i$, $cstart$, $csize$)}
\label{algo:chunk_init}
\begin{algorithmic}[1]
\For {$ k \leftarrow 0$ \textbf{to} $n-1$}
\State{$y$[$k$].$v$[$0$] $\leftarrow$ $a$[$k$]}
\State{$y$[$k$].$v$[$1$] $\leftarrow 0.0 $}
\If{ ($k = i$)}
\State{$y$[$k$].$v$[$1$] $\leftarrow 1.0$ }
%\Else
%\State{$y$[$k$].$v$[$1$] $\leftarrow 0.0 $}
\EndIf
\For {$ l \leftarrow 2$ \textbf{to}  $csize+1$}
\State {$y$[$k$].$v$[$l$] $\leftarrow 0.0$ }
\EndFor
\If {($k \geq cstart$) {\textbf and}  ($k < (cstart + csize)$}  
\State{$y$[$k$].$v$[$k-cstart+2$] $\leftarrow 1.0$ }
%\Else 
%\State{$y$[$k$].$v$[$l$] $\leftarrow 0.0$ }
\EndIf
\For {$ l \leftarrow csize+2$ \textbf{to} $2*csize + 1$}
\State{$y$[$k$].$v$[$l$] $\leftarrow$ 0.0 }
\EndFor
\EndFor
\end{algorithmic}
\end{algorithm}

\begin{algorithm}[ht]
\small
\caption{\textsc{CHUNK-HESS}($f$, $n$, $a$[$n$], $H$[$n$][$n$], $csize$)}
\label{algo:chunk-hess}
\begin{algorithmic}[1]
\State {$hDual \langle csize \rangle$  $y$[$n$], $temp$}
\State {$nchunk \leftarrow n/csize$}
\For {$ i \leftarrow 0$ \textbf{to} $n-1$}
\For {$ j \leftarrow 0$ \textbf{to} $nchunk-1$}
\State {$cstart \leftarrow  j * csize $} 
\State {{\textsc CHUNK-INIT}($n$, $y$, $a$, $i$, $cstart$, $csize$)}
\State {$temp \leftarrow f \langle hDual \langle csize \rangle \rangle (y)$}

\For {$ l \leftarrow 0$ \textbf{to} $csize-1$}
\State {$H[i][cstart + l]\leftarrow  temp.v[csize+2+l] $}
\EndFor
\EndFor
\EndFor
\end{algorithmic}
\end{algorithm}

\begin{algorithm}[ht]
\small
\caption{\textsc{SCHUNK-HESS}($f$, $n$, $a$[$n$], $H$[$n$][$n$], $csize$)}
\label{algo:schunk-hess}
\begin{algorithmic}[1]
\State {$hDual \langle csize \rangle$  $y$[$n$], $temp$}
\State {$nchunk \leftarrow n/csize$}
\For {$ i \leftarrow 0$ \textbf{to} $n-1$}
\State {$startchunk \leftarrow i/csize$}
\For {$ j \leftarrow startchunk$ \textbf{to} $nchunk-1$}
\State {$cstart \leftarrow  j * csize $} 
\State {{\textsc CHUNK-INIT}($n$, $y$, $a$, $i$, $cstart$, $csize$)}
\State {$temp \leftarrow f \langle hDual \langle csize \rangle \rangle (y)$}

\For {$ l \leftarrow 0$ \textbf{to} $csize-1$}
\State {$H[i][cstart + l]\leftarrow  temp.v[csize+2+l] $}
\EndFor
\EndFor
\EndFor
\For {$ i \leftarrow csize$ \textbf{to} $n-1$}
\State {$endindex \leftarrow (i/csize)*csize$}
\For {$ j \leftarrow 0 $ \textbf{to} $endindex$}
\State {$H[i][j] \leftarrow H[j][i] $} 
\EndFor
\EndFor
\end{algorithmic}
\end{algorithm}

\subsection{Hessian-Vector Product} The ideas mentioned in previous subsection apply to computation of the Hessian-Vector product as well. An important thing to note is that when computing the Hessian-Vector product $r = Hv$, the entry $H(i,j)$ of the Hessian contributes only to the computation of $r(i)$ and no other component. Hence, we can compute $H(i,j)$, update $r(i)$ by adding $H(i,j)*v(j)$ to it and then discard $H(i,j)$ as it is not needed any further in any other updates. Similar argument holds when Hessian is being computed in chunks. A chunk $j,j+1,\ldots j+ csize -1$ of the $i^{th}$ row of the Hessian contributes only to $r(i)$. We can compute this chunk, update $r(i)$ by adding $H(i,j)*v(j)+H(i,j+1)*v(j+1)+...H(i,j+csize-1)*v[j+csize-1]$ to it and then discard the computed chunk as it is no further needed.

If we were exploiting the symmetry of the Hessian, we will compute only the ``upper triangle" of the Hessian. In this case when $H(i,j)$ is computed we will update $r(i)$ by adding $H(i,j)*v(j)$ as before but additionally update $r(j)$ by adding $H(i,j)*v(i) = H(j,i)*v(i)$ to it. In this way, we make the correct updates but avoid computing all $H(j,i)$ for $j < i$. Once again $H(i,j)$ can be discarded after $r(i)$ and $r(j)$ have been updated. Note that if $i$ is same as $j$ then we need to make sure that we do not update $r(i)$ (which is same as $r(j)$) twice. This idea can be further extended to chunked hessian-vector product calculation that makes use of symmetry. Algorithm to compute Hessian-Vector Products in a chunked fashion are presented in section VI.

\section{CHESSFAD Library Implementation}
\label{sec:CHESSFAD-Implementation}
CHESSFAD is implemented as a standalone lightweight C++ library. It is templated on the number of second-order derivatives to be computed at a time ($csize$). It is geared towards efficient computation of the Hessian and Hessian-vector products although it can computes the Jacobian while computing the Hessian. 

CHESSFAD defines the class $hDual$. Each $hDual$ keeps an array $v$ of $2*(csize+1)$ real numbers. $v$[$0$] keeps the value of the function. $v$[$1$] keeps the first-order derivative of the function w.r.t to $x_i$ when second order derivatives in row $i$ of the Hessian are being computed. 
The next $csize$ components of $v$ namely $v$[$2$]...$v$[$csize+1$] keep the first-order derivatives of the function w.r.t to the variables in the chunk. More precisely, if the current chunk starts at index $j$, they will keep the values $\frac{\partial f}{\partial x_j}, \ldots \frac{\partial f}{\partial x_{j+ csize -1}}$. 
Finally, the next $csize$ components, namely $v$[$csize+2$] $\ldots$ $v$[$2*csize+1$], keep the values for the the second-order derivatives
$ \frac{\partial^2 f}{\partial x_i \partial x_j} \ldots \frac{\partial^2 f}{\partial x_i \partial{x_{j+ csize -1}}} $.

We then define overload operators for all arithmetic operations of interest. This includes basic operations like $+,-,*, {\mbox {and }}/$ but also include comparison operations like $ <, >, \geq, {\mbox {and }} \leq$ and operations like $abs$, $sin$, $cos$ and $exp$. The overload operators are defined in a way so as to propagate the correct values of the function and all the derivatives.

To give a better idea of how this is done we include below code snippets from the library that illustrate the structure of class $hDual$ and implementation of the $+$ and  $*$ operator for $hDual$s. 

\begin{figure}
 \centering 
\begin{minted}[frame=lines, fontsize=\footnotesize]
{c++}

template <int csize>
class hDual {
public:
  double v[2*csize + 2]; 

hDual operator+(const hDual &op1, const hDual &op2)
{ 
  hDual result;
  for (int i=0; i <= (2*csize + 1); i++ ) { 
    result.v[i] = op1.v[i] + op2.v[i];
  }
  return result;
}

hDual operator+(const double &op1, const hDual &op2)
{
  hDual result;
  result.v[0] = op1 + op2.v[0];
  for (int i=1; i <= (2*csize + 1); i++ ) { 
    result.v[i] = op2.v[i];
  }
  return result;
}

hDual operator+(const hDual &op1, const double &op2)
{
  hDual result;
  result.v[0] = op1.v[0] + op2;
  for (int i=1; i <= (2*csize + 1); i++ ) { 
    result.v[i] = op1.v[i];
  }
  return (result);
}

hDual operator*(const hDual &op1, const hDual &op2)
{
  hDual result;
  result.v[0] = op1.v[0]*op2.v[0];
  //compute the first derivatives:  
  //duv/dxj = udv/dxj +  vdu/dxj
  for (int i=1; i<=csize+1; i++)
    result.v[i] = op1.v[0]*op2.v[i] 
                + op2.v[0]*op1.v[i]; 

  //compute a chunk of ith row of the Hessian. 
  for (int j=2; j<=csize+1; j++){
    result.v[csize+j] = 
        op1.v[0]*op2.v[csize+j] +
        op1.v[1]*op2.v[j] +
        op2.v[1]*op1.v[j] +
        op2.v[0]*op1.v[csize+j]; 
  }
  return result;
}

hDual operator*(const hDual &op1, double &op2)
{
  hDual result;
  for (int i=0; i <= (2*csize + 1); i++)
    result.v[i] = op2*op1.v[i];
  return result;
}

hDual operator*(double &op1, const hDual &op2)
{
  hDual result;
  for (int i=0; i <= (2*csize + 1); i++){
    result.v[i] = op1*op2.v[i];
  }
  return result;
}


\end{minted}
 \caption{Templated C++ Class {\tt hDual} with overload operators + and *.}  
\label{fig:hdual}
\end{figure}

\section{Analysis of Scalar Operations Needed for computing the Hessian}
\label{sec:Analysis}

We provide mathematical analysis for the number of scalar operations needed for computing the Hessian and Hessian-vector product for a real-valued function $f$ of $n$ variables $f(x_1,x_2,\ldots x_n)$ in our methods. We are interested in how this varies with chunk size $csize$ and $n$ as well as number of operations in $f$. For illustration purposes, we assume that $f$ uses only scalar addition and multiplication operations. Similar analysis can be carried out when other operations are involved. For simplicity, we also assume that $n$ is exactly divisible by $csize$. 

We begin by noting that an addition operation of $hDual\langle csize \rangle$s requires $2*csize + 2$ scalar additions (and 0 scalar multiplications). Similarly, a multiplication of two {\tt hdual<CHUNK>}s requires ($6*csize +3$) scalar multiplications and $4*csize$ additions. (see code snippet in previous section). 

Let us assume that the function $f$ variables requires $M$ multiplication and $A$ addition operations. Then, if we compute the Hessian of $f$ using Algorithm 4 ({\textsc CHUNK-HESS}), we will call the templated function $f$ exactly $n*n/csize$ times. Each of these calls will require $M$ multiplications and $A$ additions of {$hDual\langle csize \rangle$}s. For $M$ $hDual\langle csize\rangle$ multiplications, we will require a total of $(n^2/csize )*(6*csize + 3)*M$
= $n^2(6 + 3/csize)*M$ scalar multiplications and $(n^2/csize)*4*csize*M$ = $n^2*4*M$ scalar additions. Similarly, for $A$ $hDual\langle csize\rangle$ additions, we will require $(n^2/csize)*(2*csize+2)*A$ = $(2 + 2/csize)*n^2*A$ scalar additions. To summarize, we will need a total of $(6 + 3/csize)*n^2*M$ scalar multiplications and $4*n^2*M + (2 + 2/csize)*n^2A$ additions. Note that in this simple way of doing the Hessian calculation (which has $n^2$ elements), the number of scalar multiplication operations gets multiplied by a factor of $c(csize)*n^2$ where $c(csize) = 6 + 3/csize$ is bounded below by $6$ and is a monotone decreasing function of $csize$. Hence choosing larger value of $csize$  is better for reducing the scalar multiplications. The number of scalar additions also decreases as $csize$ increases. 

Next, we present an analysis of number of scalar operations needed if the algorithm exploiting symmetry (\textsc SCHUNK-HESSIAN) was used for computing the Hessian of $f$. The key here is that Hessian for many of the chunks (those completely in the bottom half) are not computed but rather updated using symmetry which provides substantial savings. More precisely, the number of chunks for which the Hessian is computed is:\\
\\
$csize*(n/csize) + csize*(n/csize -1) + \ldots + csize*1$\\
= $csize*(1+2+\ldots n/csize)$\\
= $csize*(n/csize)*(n/csize +1)/2$\\ 
= $n*(n/csize +1)/2$.\\
If $csize$ is much less than $n$, this is about  half of the number of chunks for which the Hessian is computed in the first method ($n^2/csize$) as expected. Note that if $csize = 1$ then this number is $n(n+1)/2$ and if $csize = n$ this equals $n$. The number of scalar multiplications needed then is:\\
\\
$(n*(n/csize +1)/2)*(6*csize +3)*M$ \\
= $(3/2)*n*(n/csize + 1)(2*csize+1)*M$ \\
= $(3/2)*n*(2*n + 2*csize + n/csize + 1)*M$.\\

Interestingly, for fixed $n$ this quantity is minimized at $csize = \sqrt{n/2}$ which tells us that for minimizing the number of scalar multiplications one should select $csize$ as close to that as possible. In particular, for $n=2^{2k+1}$ this choice is $2^k$. 
Similar analysis can be carried out when other operations are involved. 

We do note that the actual performance of an algorithm implementation is not simply dependent on the number of arithmetic operations needed. Other factors like memory usage are important and could play quite a significant role, especially in the case of GPU implementations.

% Table TABLE gives a summary of number of scalar operations required for each {\tt hdual<CHUNK>} operation to aid such analysis.

\section{Hessian-Vector Product Implementation}
\label{sec:HVP-Implementation}
A high-level description of the sequential Hessian-vector product computation for a single instance using chunked Hessian is outlined in Algorithm~\ref{algo:chunked-hessian-vector}.  The inputs to the algorithm are the function $f$, the number of variables $n$ for $f$, the array $a$ of $n$ variables which is the $n$-dimensional point at which the Hessian of $f$ is to be computed, the input vector $in$ of size $n$ that needs to be multiplied by the Hessian, and chunk size $csize$. The output $out$ is a vector of size $n$.  The computation consists of a loop over $n$, where we compute a row of the Hessian matrix in each iteration, multiply by the corresponding elements of the input vector, and aggregate the result. The {\em templated} function  $f$ on line~8 performs the Hessian computation. In the for loop in line~9, we aggregate the result of the multiplication of a row of Hessian with the input vector.  

\begin{algorithm}[ht]
\small
\caption{\textsc{CHESS-VEC}($f$, $n$, $a[n]$, $in[n]$, $out[n]$, $csize$)}
\label{algo:chunked-hessian-vector}
\begin{algorithmic}[1]
\State {$hDual \langle csize \rangle$  $y[n]$, $temp$}
\State {$nchunk \leftarrow n/csize $}
\For {$ i \leftarrow 0$ \textbf{to} $n-1$}
\State{$res \leftarrow 0$}
\For {$ j \leftarrow 0$ \textbf{to} $nchunk-1$}
\State {$cstart \leftarrow  j * csize $} 
\State {{\textsc{CHUNK-INIT}}$(y, a, n, i, j, csize)$}
\State {$temp\leftarrow f\langle hDual\langle csize \rangle \rangle (y)$}

\For {$ l \leftarrow 0$ \textbf{to} $csize-1$}
\State {$res\leftarrow  res + temp.v[csize+2+l] *in[cstart+l]$}
\EndFor
\EndFor
\State{$out[i] \leftarrow res$}
\EndFor
\end{algorithmic}
\end{algorithm}

Once again, one can exploit the symmetry of the Hessian to reduce the number of calculations. Algorithm~\ref{algo:symmetric-chunked-hessian-vector} accomplishes that.

\begin{algorithm}[ht]
\small
\caption{\textsc{SC-HESS-VEC}($f$, $n$, $a[n]$, $in[n]$, $out[n]$, $csize$)}
\label{algo:symmetric-chunked-hessian-vector}
\begin{algorithmic}[1]
\State {$hDual \langle csize \rangle$  $y[n]$, $temp$}
\State {$nchunk \leftarrow n/csize $}
\State {$double$ $res[n] \leftarrow 0.0$};
\For {$i \leftarrow 0$ \textbf{to} $n-1$}
\State {$scn \leftarrow i/csize$}
\For {$cn \leftarrow scn$ \textbf{to} $nchunk-1$}
\State{$cstart \leftarrow cn*csize$}
\State{$s \leftarrow cstart$}
\State {{\textsc{CHUNK-INIT}}$(n,y, a, n, i, cstart, csize)$}
\State {$temp\leftarrow f\langle hDual\langle csize \rangle \rangle (y)$}

\For {$ l \leftarrow csize+2$ \textbf{to} $2*csize-1$}
\State {$res[i] \leftarrow  res[i] + temp.v[l] *in[s]$}
\If{$(cn > scn)$}
\State{$res[s] \leftarrow res[s] + temp.v[l]*in[i]$}
\EndIf
\State {$s \leftarrow s+1$}
\EndFor
\EndFor
%\State{$out[i] \leftarrow res$}
\EndFor

\end{algorithmic}
\end{algorithm}

GPU implementation launches several thousands of instances concurrently. The first level of parallelism, $L0$, is over multiple instances: the evaluation of Hessian-vector product over multiple $n$-dimensional data points, see Algorithm~\ref{algo:gpu-kernel-l0}. The input and output arrays are of size $n \times m$, where $m$ is the number of instances launched by the GPU. In this algorithm, a thread is assigned to process an instance. In line~2, we set the thread $id$, which is then used to access data relevant to its instance. In the second level of parallelism, $L1$, we additionally parallelize a single instance over $n$ rows of the Hessian matrix. In terms of implementation, it implies spreading the for loop in line~4 of Algorithm~\ref{algo:gpu-kernel-l0} over $n$ threads. Note that for $L1$ parallelism we assign $n$ threads to process an instance. Algorithm~\ref{algo:gpu-kernel-l1}  gives an overview of the implementation of $L1$. In the third level of parallelism, $L2$, we further partition the computation of a Hessian row into chunks, where different chunks can be computed concurrently. The loop in line~6 of Algorithm~\ref{algo:gpu-kernel-l0} is spread over $nchunk$ (number of chunks) threads (similar to Algorithm~\ref{algo:gpu-kernel-l1}).  Note that for $L2$ parallelism we assign $n \times nchunk$ threads to process an instance. In the  $L2$  implementation,  the multiplication of a row of Hessian with the vector is distributed across multiple threads. The partial results that are spread across threads need to be aggregated, which is accomplished using the shared memory. The code segment that illustrates the $L2$ implementation is shown in Figure~\ref{fig:cuda-code-l2}. 

\begin{algorithm}[ht]
\small
\caption{\textsc{L0-HESS-VEC}$(f, n, a[n*m], in[n*m], out[n*m], csize)$}
\label{algo:gpu-kernel-l0}
\begin{algorithmic}[1]
\State {$hDual \langle  csize \rangle$  $y[n]$, $temp$}
\State{$id \leftarrow blockIdx.x * blockDim.x + threadIdx.x$}
\State {$ nchunk \leftarrow  n/csize$}
\For {$ i \leftarrow 0$ \textbf{to} $ n-1$}
\State{$res \leftarrow 0$}
\For {$ j \leftarrow 0$ \textbf{to} $ nchunk-1$}
\State {$ cstart \leftarrow  j*  csize $} 
\State {{\textsc{CHUNK-INIT}} $(y, \&a[id*n], n, i, j, csize)$}
\State {$temp \leftarrow  f \langle  hDual \langle csize \rangle \rangle (y)$}

\For {$ l \leftarrow 0$ \textbf{to} $ csize-1$}
\State {$\mathrm{res} \leftarrow  \mathrm{res} + \mathrm{temp.v[csize+2+l]} * \mathrm{in[cstart+l]}$}
\EndFor
\EndFor
\State{$out[id*n+i] \leftarrow res$}
\EndFor
\end{algorithmic}
\end{algorithm}

\begin{algorithm}[ht]
\small
\caption{\textsc{L1-HESS-VEC}$(f, n, a[n*m], in[n*m], out[n*m], csize)$}
\label{algo:gpu-kernel-l1}
\begin{algorithmic}[1]
\State {$ hDual \langle  csize \rangle$  $y[n]$, $temp$}
\State{$id \leftarrow$ blockIdx.x * blockDim.x + threadIdx.x}
\State{$eid \leftarrow id/n$}
\State{$i \leftarrow id\%n$}
\State {$ nchunk \leftarrow  n/csize$}
\State{$res \leftarrow 0.0$}
\For {$ j \leftarrow 0$ \textbf{to} $ nchunk-1$}
\State {$ cstart \leftarrow  j *  csize $} 
\State {{\textsc{CHUNK-INIT}}$(y, \&a[eid*n], n, i, j, csize)$}
\State {$temp \leftarrow  f \langle  hDual \langle csize \rangle \rangle (y)$}

\For {$ l \leftarrow 0$ \textbf{to} $csize-1$}
\State {$res \leftarrow  res + temp.v[csize+2+l] * in[cstart+l]$}
\EndFor

\State{$out[eid*n+i] \leftarrow res$}
\EndFor
\end{algorithmic}
\end{algorithm}

\begin{figure}
 \centering 
\begin{minted}[frame=lines, fontsize=\footnotesize]
{c++}

  // NV: number of variables
  // NCHUNK: number of chunks
  const int local_eid = threadIdx.x / (NV * NCHUNK);
  const int eid = NETBLK * blockIdx.x + local_eid; 
  const int tid = threadIdx.x % (NV * NCHUNK);
  const int i = tid / NCHUNK;
  const int j = tid % NCHUNK;

  INITIALIZE<NV>(y, &x[eid * NV], i, j, CHUNK);
  temp1 = rosenbrock<hDual, NV>(y);
  chunkstart = j * CHUNK;
  for (int l = CHUNK + 2; l <= 2 * CHUNK + 1; l++)
  {
    res = res + temp1.v[l] * vec[eid * NV + 
                  chunkstart + l - CHUNK - 2];
  }
  // save partial results in shared memory  
  sprod[local_eid][i][j] = res;
  __syncthreads();
  // accumulate results in the shared memory
  if (tid < nvar)
  {
    double res = 0.0;
    for (int k = 0; k < NCHUNK; k++)
    {
      res = res + sprod[local_eid][tid][k];
    }
    z[tid + eid * NV] = res;
  }
   

\end{minted}
 \caption{CUDA code segment for $L2$ implementation.}  
\label{fig:cuda-code-l2}
\end{figure}

% \begin{algorithm}[ht]
% \small
% \caption{\textsc{GPU-KERNEL}(n, a[n*npoints], in[n*npoints], out[n*npoints], csize, rfun)}
% \label{algo:gpu-kernel}

% \begin{algorithmic}[1]
% \State {$\mathrm{hDual} \langle \mathrm{csize} \rangle$  y[n], temp}
% \State{id $\leftarrow$ blockIdx.x * blockDim.x + threadIdx.x}
% \State{eid $\leftarrow$ id / n}

% \State{i $\leftarrow$ id \% NV}
% \State {$\mathrm{nchunk} \leftarrow \mathrm{n/csize}$}
% \For {$ i \leftarrow 0$ \textbf{to} $ n-1$}
% \For {$ j \leftarrow 0$ \textbf{to} $\mathrm{nchunk}-1$}
% \State {$\mathrm{cstart} \leftarrow  \mathrm{j} * \mathrm{csize} $} 
% \State {$\mathrm{INITIALIZE}$(y, \&a[eid * n], n, i, j, csize)}
% \State {$\mathrm{temp} \leftarrow \mathrm{rfun} \langle \mathrm{hDual} \langle csize \rangle \rangle (y)$}

% \For {$ l \leftarrow 0$ \textbf{to} $\mathrm{csize}-1$}
% \State {$\mathrm{res} \leftarrow  \mathrm{res} + \mathrm{temp.v[csize+2+l]} * \mathrm{in[eid * n + cstart+l]}$}
% \EndFor
% \EndFor
% \State{out[i + eid * n] $\leftarrow$ res}
% \EndFor
% \end{algorithmic}
% \end{algorithm}

\section{Experimental Results}
\label{sec:Experimental-Results}
We are interested in how the execution time grows as we increase the number of variables. We expect that for values of $n$ greater than some threshold, the execution time, which depends on the nature of $func$ and the fact that the Hessian-vector product has a quadratic run-time complexity, will become prohibitively large. We experimented with three standard multivariate function families, Rosenbrock, Ackley, and Fletcher-Powell\cite{Fletcher1963ARC}. We compared the performance of CHESSFAD with two other C++ header-based libraries HAD\cite{}, and {\tt autodiff} \cite{autodiff}.  As mentioned earlier these two libraries are only supported on CPUs and do not work on GPUs.

We evaluated execution time trends on the CPU as well as on the GPU. We did the sequential implementation on the Intel Core-i7 CPU, and used GNU g++ compiler. The GPU implementation was done on NVIDIA A100 GPU with 80GB HBM2e memory using CUDA APIs. We collected execution time results for $1K$ data points on CPU and $0.5M$ data points on GPU. The GPU execution time reported in this paper is only the kernel time and does not include time to transfer data between CPU and GPU.  Note that our objective of CPU implementation was to evaluate the performance behavior with an increase in the number of variables, and not to compare the performance with GPU\footnote{ For comparing CPU and GPU performance it is important to have a parallel implementation on CPU as well to exploit all the cores of the CPU, which we do not have}. Figures~\ref{fig1z}-\ref{fig3z} show the execution time trends for three implementations as we increase the number of variables for the three test functions.  We observe that for small values of $n$,   HAD performs poorly, but outperforms {\tt autodiff} and CHESSFAD for larger values of $n$.  The crossover point for the Rosenbrock function happens at close to $n=14$ (See Figure~\ref{fig4z}),  and for Ackley it happens at $n=10$ (See Figure~\ref{fig5z}).  For values of $n$ less than the crossover value, CHESSFAD and {\tt autodiff} perform better. However, for the Fletcher-Powell function HAD performs poorly for all values of $n$ (see Figure~\ref{fig6z}). When we compare CHESSFAD and {\tt autodiff}, we found CHESSFAD to perform better,  on average $20\%$  for Rosebrock, $5\%$ for Ackley, and $49\%$ for Fletcher-Powell (See Figure~\ref{fig7z}).

The GPU execution time trends for three test functions and three implementations are summarized in Figures~\ref{fig8z}-\ref{fig10z}. The trend for all three functions is similar, the execution time grows as the increase in number of variables.  We also observe that the $L2$ implementation performs better than the other two implementations as the number of variables increases. Our experiments indicated that the growth of CPU execution time with the number of variables is slower than the GPU execution time growth. This can be observed in Table~\ref{fig:table_rosenbrock}, Table~\ref{fig:table_ackley}, Table~\ref{fig:table_fletcher}, and Figure~\ref{fig11z}, where we summarize the behavior of normalized GPU execution time that is the GPU execution time per data point divided by the CPU execution time per data point, also referred to as speedup. Note that the speedup is high at $n=2$, but it starts decreasing with the number of variables.  This is true with all three functions. Despite this efficiency loss for GPU implementations, the timing results indicate that parallel implementations of CHESSFAD on GPU enable Hessian-vector computation of a function with many more variables compared to the sequential implementation. For example, the sequential execution time for the Hessian-vector product for two variables is approximately enough to compute the Hessian-vector product for 16 variables on GPU for all three functions. 
 
\begin{figure}[htbp]
\centerline{\includegraphics[scale=0.4]{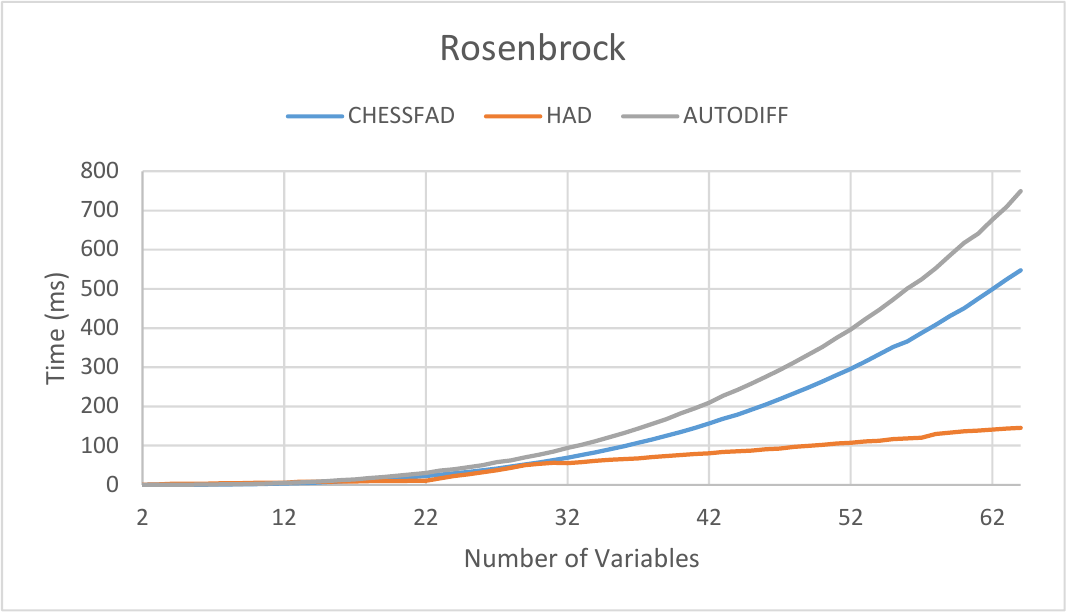}}
\caption{Execution time trend for sequential implementations for Rosenbrock function.}
\label{fig2z}
\end{figure}

\begin{figure}[htbp]
\centerline{\includegraphics[scale=0.4]{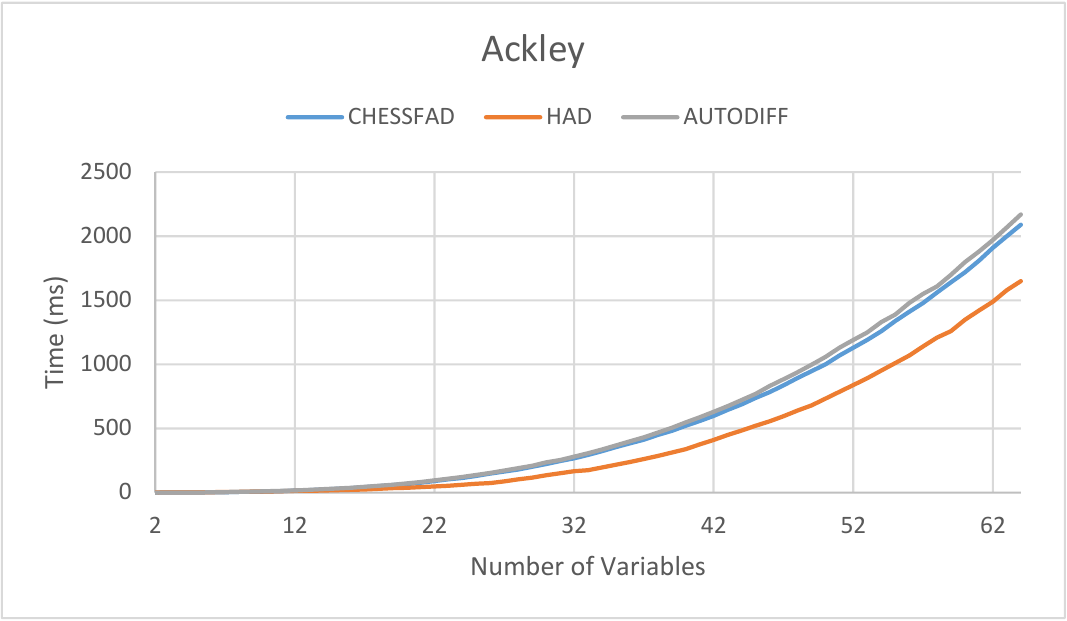}}
\caption{Execution time trend for sequential implementations for Ackley function.}
\label{fig4z}
\end{figure}

\begin{figure}[htbp]
\centerline{\includegraphics[scale=0.4]{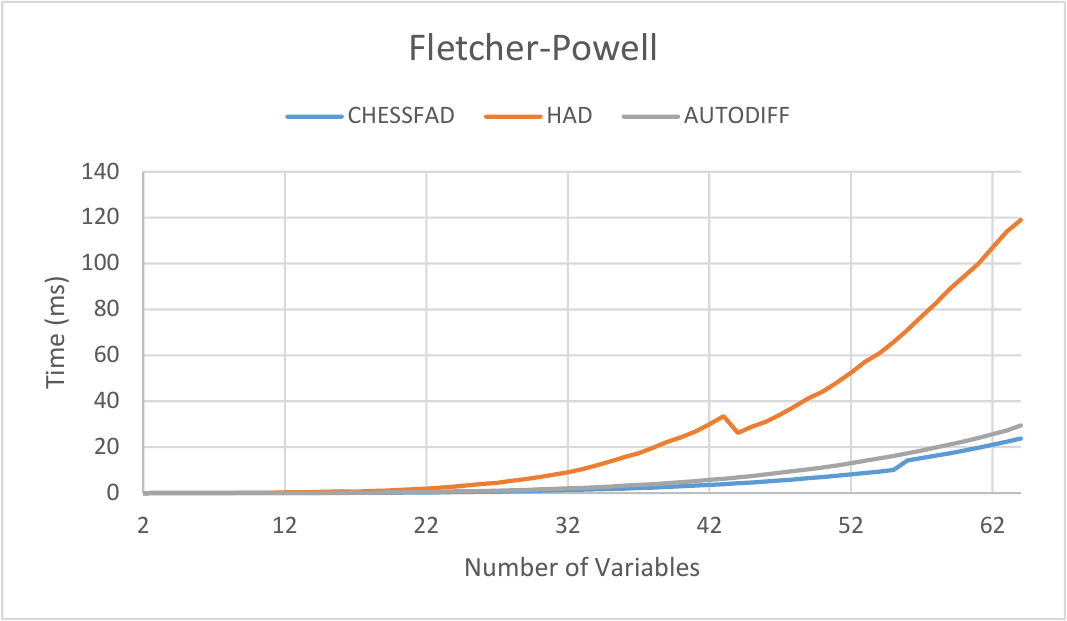}}
\caption{Execution time trend for sequential implementations for Fletcher-Powell function.}
\label{fig5z}
\end{figure}

%-----------------------------------------------------

\begin{figure}[htbp]
\centerline{\includegraphics[scale=0.4]{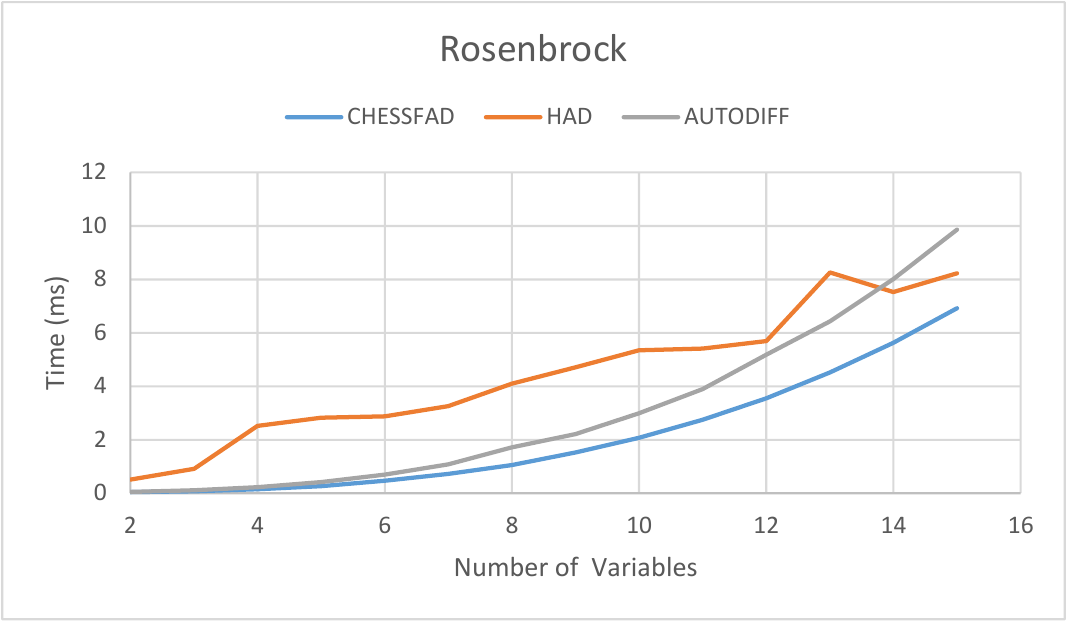}}
\caption{Execution time trend for sequential implementations for Rosenbrock function for small values of $n$.}
\label{fig1z}
\end{figure}

\begin{figure}[htbp]
\centerline{\includegraphics[scale=0.4]{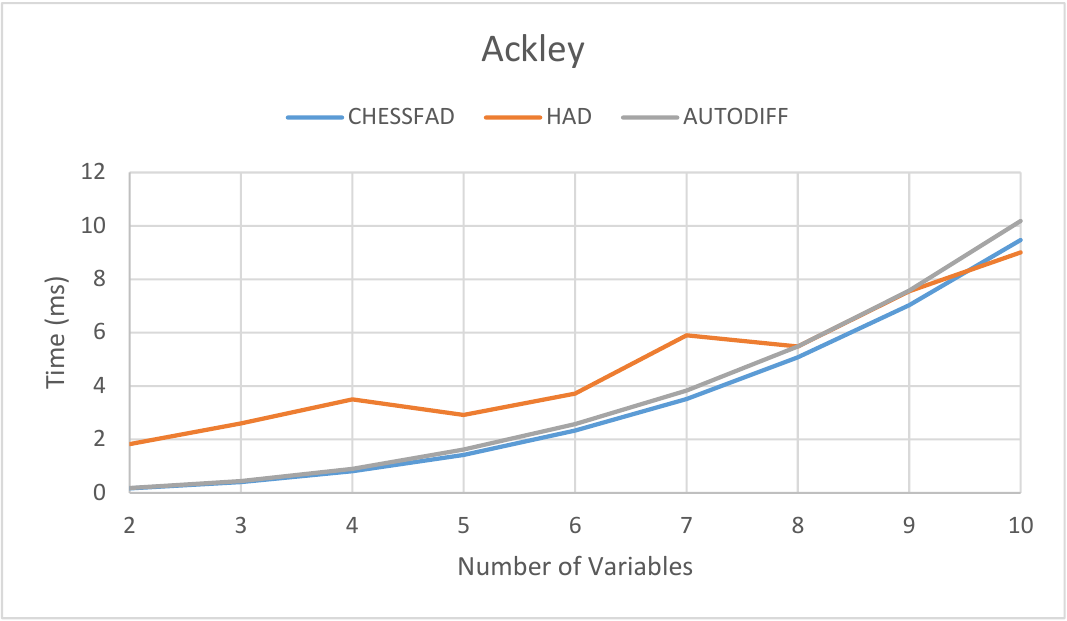}}
\caption{Execution time trend for sequential implementations for Ackley function for small values of $n$.}
\label{fig3z}
\end{figure}

\begin{figure}[htbp]
\centerline{\includegraphics[scale=0.4]{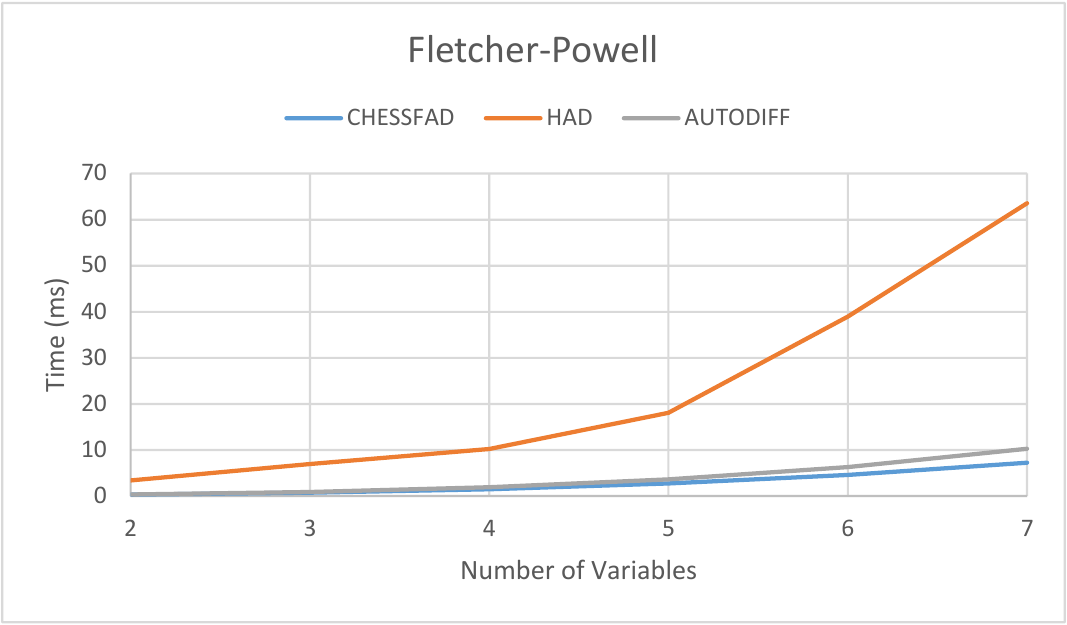}}
\caption{Execution time trend for sequential implementations for Fletcher-Powell function for small values of $n$.}
\label{fig6z}
\end{figure}

\begin{figure}[htbp]
\centerline{\includegraphics[scale=0.4]{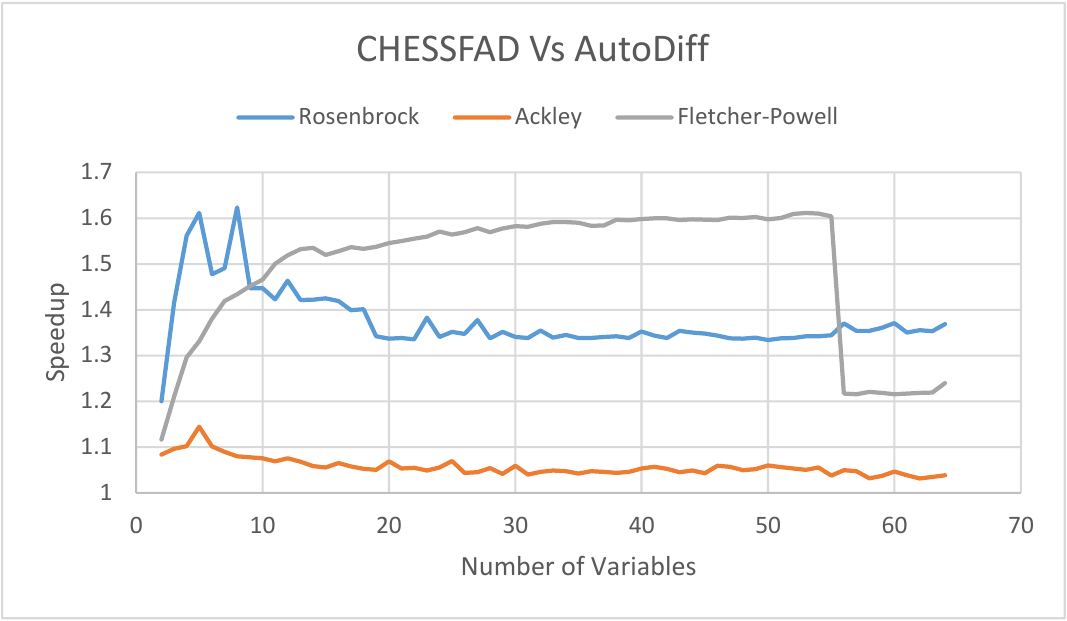}}
\caption{Comparison of CHESSFAD with {\tt autodiff}. }
\label{fig7z}
\end{figure}

%-----------------------------------------------------

\begin{figure}[htbp]
\centerline{\includegraphics[scale=0.4]{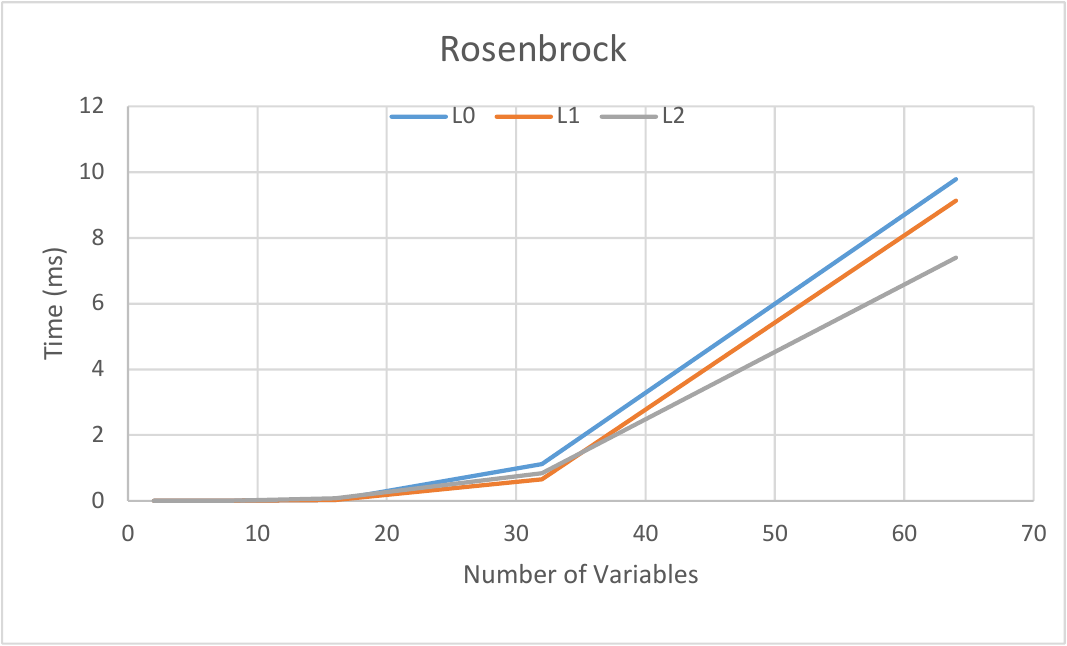}}
\caption{Execution time trend for GPU implementations for Rosenbrock function.}
\label{fig8z}
\end{figure}

\begin{figure}[htbp]
\centerline{\includegraphics[scale=0.4]{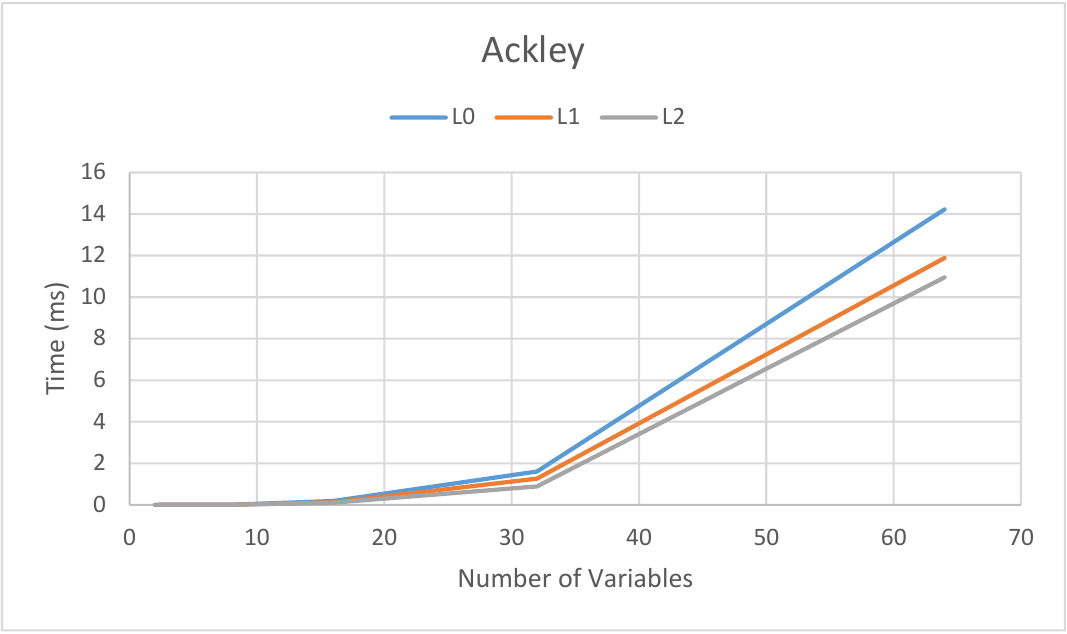}}
\caption{Execution time trend for GPU implementations for Ackley function.}
\label{fig9z}
\end{figure}

\begin{figure}[htbp]
\centerline{\includegraphics[scale=0.4]{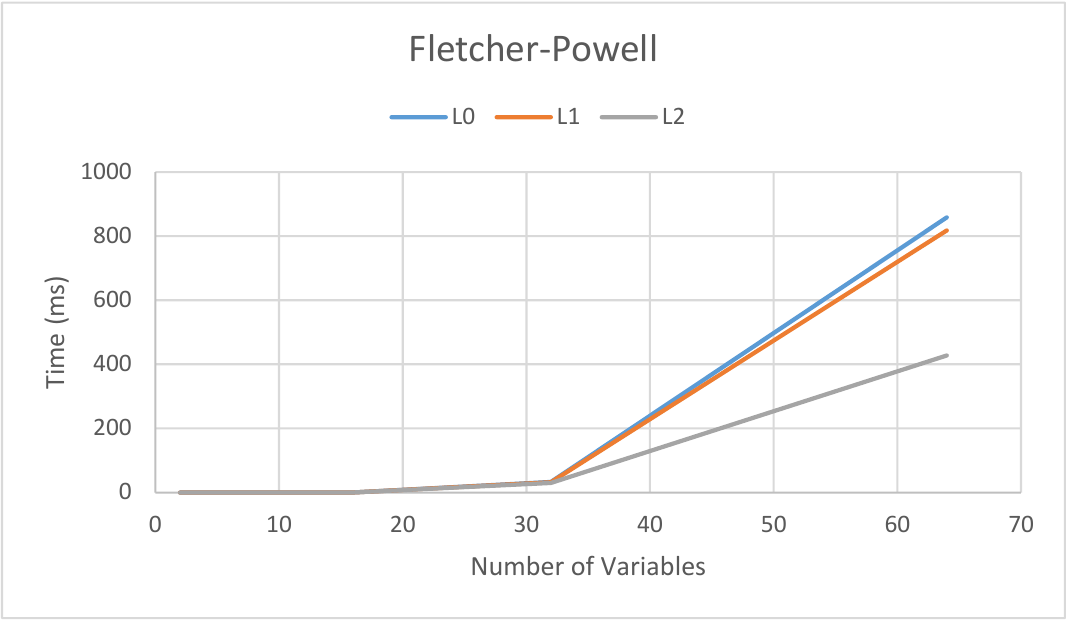}}
\caption{Execution time trend for GPU implementations for Fletcher-Powell function.}
\label{fig10z}
\end{figure}

\begin{table}[ht]
\caption{Normalized GPU execution time for Rosenbrock function.}
\centerline{\includegraphics[scale=0.9]{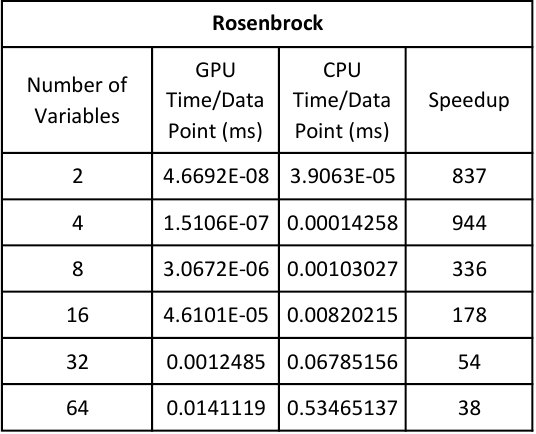}}
\label{fig:table_rosenbrock}
\end{table}

\begin{table}[ht]
\caption{Normalized GPU execution time for Ackley function.}
\centerline{\includegraphics[scale=0.9]{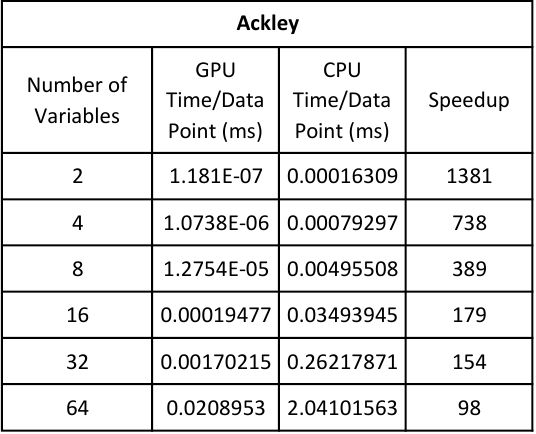}}
\label{fig:table_ackley}
\end{table}

\begin{table}[ht]
\caption{Normalized GPU execution time for Fletcher-Powell function.}
\centerline{\includegraphics[scale=0.9]{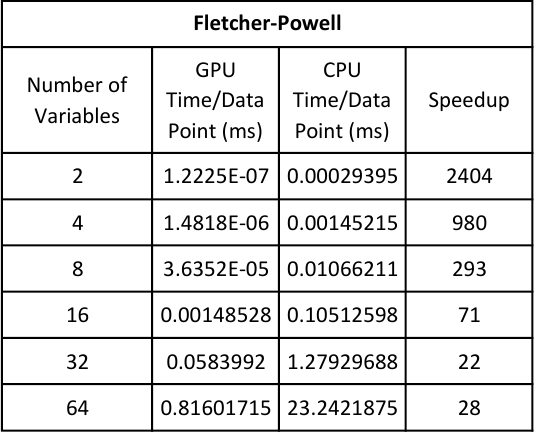}}
\label{fig:table_fletcher}
\end{table}

\begin{figure}[htbp]
\centerline{\includegraphics[scale=0.4]{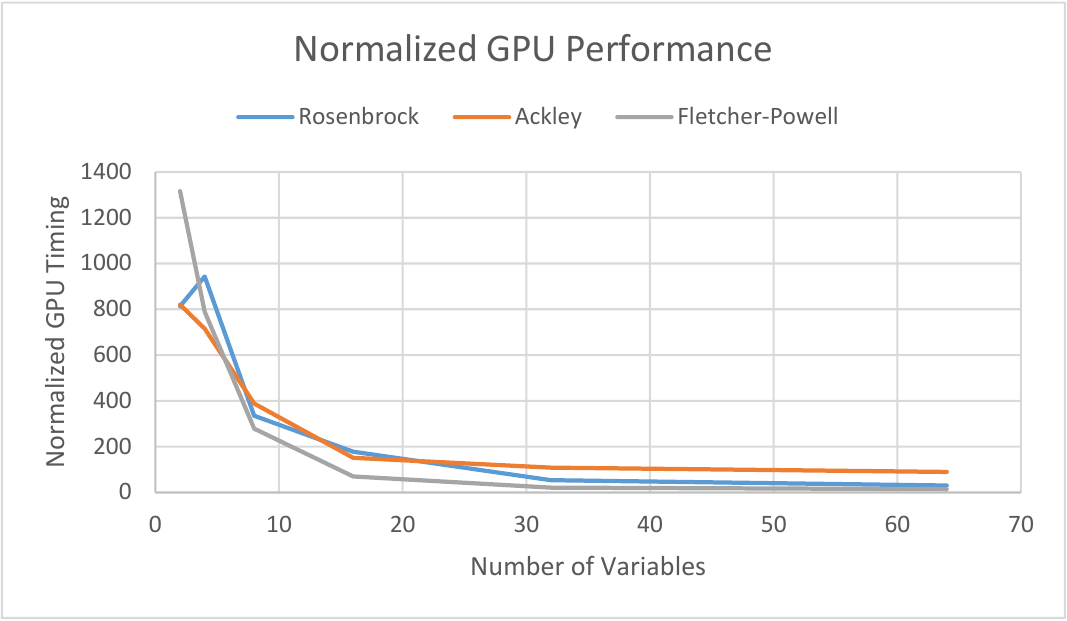}}
\caption{Normalized GPU Performance. }
\label{fig11z}
\end{figure}

\section{Conclusion}
\label{sec:Conclusion}
In this paper, we proposed an automatic differentiation approach, CHESSFAD,  for finding second-order derivatives of a function written in C++. CHESSFAD is implemented as a lightweight header-based C++ library that works both for CPUs and GPUs. Our approach to Hessian computation exposes parallelism at different levels. The Level~0 parallelism is across multiple data instances allowing concurrent computation of Hessian of a function at multiple data points. In Level~1, rows of the Hessian matrix are computed concurrently.  For Level~2, we partition the computation of a Hessian row into chunks, where different chunks can be computed concurrently. We evaluated the performance of CHESSFAD for performing a large number of concurrent, independent Hessian-Vector products on a set of standard test functions, and compared its performance to other existing header-based C++ libraries such as {\tt autodiff}. Compared to {\tt autodiff}, we found CHESSFAD to perform better,  on average $20\%$  for Rosebrock, $5\%$ for Ackley, and $49\%$ for Fletcher-Powell. We also analyze its efficiency on GPUs as the number of variables in the function grows. We demonstrated that our approach is easily parallelizable and enables us to work with Hessian of a function of a large number of variables, which was not possible in sequential implementation. 

\bibliographystyle{unsrt}
\bibliography{CHESSFADarxiv}  

\end{document}